\documentclass{amsart} 
\usepackage{graphicx}
\usepackage{sidecap}
\usepackage{amsmath}
\usepackage{amsfonts}
\usepackage{amssymb}
\usepackage{float}

\newcommand{\R}{\mathbb R}

\def\la{\label}

\newtheorem{thm}{Theorem}[section]
\newtheorem{lem}{Lemma}[section]
\newtheorem{defi}{Definition}[section]
\newtheorem{rem}{Remark}[section]

\numberwithin{equation}{section}
\numberwithin{figure}{section}

\def\bt{\begin{thm}}
\def\et{\end{thm}}
\def\bl{\begin{lem}}
\def\el{\end{lem}}
\def\bd{\begin{defi}}
\def\ed{\end{defi}}
\def\bc{\begin{cor}}
\def\ec{\end{cor}}
\def\bp{\begin{proof}}
\def\ep{\end{proof}}
\def\br{\begin{rem}}
\def\er{\end{rem}}

\begin{document}

\title[Nature of Andrews Critical Point]{Third-Order Gas-Liquid Phase Transition and the Nature of Andrews Critical Point}
\author[Ma]{Tian Ma}
\address[TM]{Department of Mathematics, Sichuan University,
Chengdu, P. R. China}

\author[Wang]{Shouhong Wang}
\address[SW]{Department of Mathematics,
Indiana University, Bloomington, IN 47405}
\email{showang@indiana.edu}

\thanks{The work was supported in part by the
Office of Naval Research and by the National Science Foundation.}


\begin{abstract} 
The main objective of this article is to study the nature of the  Andrews critical point in the gas-liquid transition in a physical-vapor transport (PVT) system. A dynamical model, consistent with the van der Waals equation near the Andrews critical point, is derived. With this model, we deduce  two physical parameters, which interact exactly at the Andrews critical point, and  which dictate the dynamic transition behavior near the Andrews critical point. In particular, it is shown that  1) the Andrews critical point is a switching point where the phase transition changes from the first order to the third order,  2) the gas-liquid co-existence curve can be extended beyond the Andrews critical point, and 3) the liquid-gas phase transition going beyond Andrews point is of the third order.  This clearly explains why it is hard to observe the gas-liquid phase transition  beyond the Andrews critical point. Furthermore, the analysis  leads naturally the introduction of  a general asymmetry principle of 
fluctuations  and the preferred transition mechanism for a thermodynamic system. 
\end{abstract}

\maketitle
\section{Introduction}
\label{sc1}
Phase transition  is one of the central problems in nonlinear sciences. Many systems have different phases, and the most commonly encountered phases are gas, liquid and solid phases. A natural system which possesses these three phases is the physical-vapor transport (PVT) system. As we know, a $PVT$ system is a system composed of one type  of molecules, and  the interaction between molecules is governed by the van
der Waals law. The molecules generally have a repulsive core and
a short-range attraction region outside the core. Such systems
have a number of phases: gas, liquid and solid, and a solid can
appear in a few phases. The most typical example of a $PVT$ system is
water. 

A $PT$-phase diagram of a typical $PVT$ system is schematically
illustrated by Figure \ref{f8.1-1}, where point $A$ is the triple point
at which the gas, liquid, and solid phases can coexist. Point $C$
is the Andrews critical point at which the gas-liquid coexistence
curve terminates \cite{reichl, stanley}.  Classical view on the termination of  the 
gas-liquid coexistence curve at the critical point amounts to saying that the system  
can go continuously from a gaseous  state 
to a liquid state without ever meeting an observable phase transition,
if we choose the right path.
\begin{figure}
  \centering
  \includegraphics[height=.23\hsize,clip]{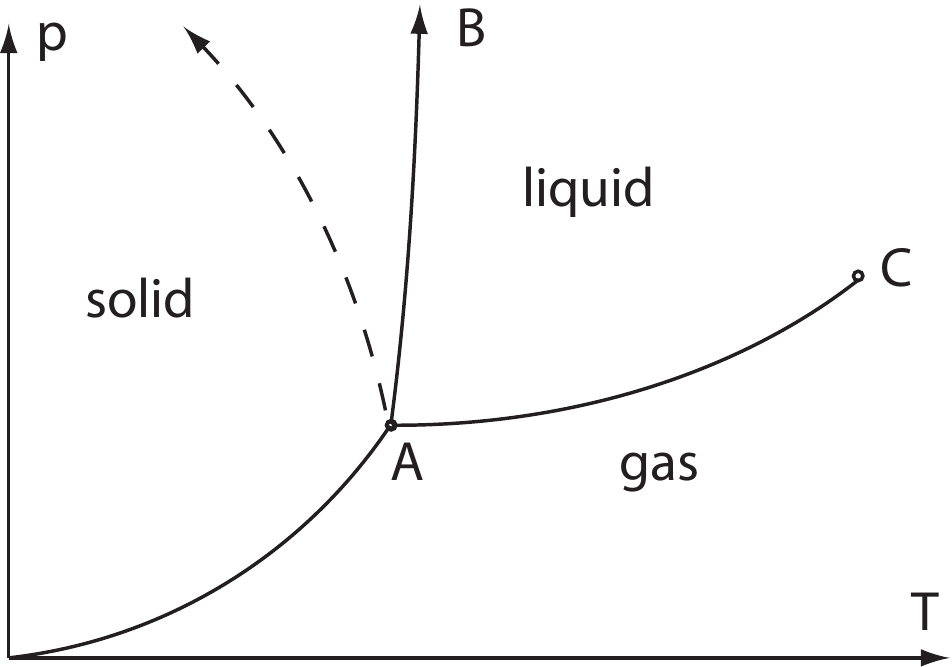}
 \caption{Coexistence curves of a typical $PVT$ system:
$A$ is the triple point, $C$ is the critical point, and the dashed
cure is a melting line with negative slope.}\la{f8.1-1}
 \end{figure}

It is, however,  still an open question why the  Andrews critical point exists  and what is the order of transition going beyond this critical point.  In \cite{MW08a}, a mathematical theory is derived to address this problem. In this article, we explore the physical implications of the mathematical theory derived in \cite{MW08a},  give a theory on the nature of the Andrews critical point, and introduce the asymmetry principle of fluctuations and the preferred transition mechanism. 

\bigskip

{\sc First,}  the modeling is  based on 1) the Landau mean field theory, 2) a  unified dynamic approach for equilibrium phase transitions, 3)  the classical phase diagram in Figure~\ref{f8.1-1}, and c) the van der Waals equation. It is worth mentioning two important aspects of the model we derived. One  is  that the new model  can be used to study liquid-solid and gas-solid transitions as well, by choosing different parameters. Second  is the consistency of the model with the van der Waals equation. 
Namely, near the Andrews  critical point $C$,  the steady state equation of the homogeneous model 
(\ref{8.11}) is exactly the van der Waals equation. This consistency gives a good validation of the mean field model. In addition,  the dynamic  approach leading to the model provides  much richer information. For example, the model (\ref{8.7}) can be used to study the heterogeneity of the system.

\bigskip

{\sc Second, } with the dynamic model at our disposal, we  introduce two new physical parameters $\lambda=\lambda(T, p)$  and $a_2=a_2(T, p)$, where the temperature $T$  and the pressure  $p$   are control parameters.  These two physical parameters determine the phase transition behavior near the Andrews point and provide the key ingredient to characterize the nature of the Andrews point.   

It is remarkable that these two parameters reproduce  the location of the 
Andrews critical point $C$, which is the same as derived by van der Waals in his classical work, although the method we use is the dynamic approach based on the Landau mean field theory, different from the one used by van der Waals.
Coincidentally,  these two parameters are the second and third-order  derivatives of the Gibbs energy at the equilibrium state $\rho_0$; see (\ref{2p}).

\bigskip

{\sc Third}, the two parameters $\lambda$ and $a_2$ correspond to two-curves in the $pT$-phase plane, and interact  exactly at the Andrews critical point $C$. Then  with the dynamic transition theory developed recently by the authors,  we deduce a theory on  the Andrews critical point $C$: 1) the critical point is a switching point where the phase transition changes from the first order with latent heat to the third order, and 2) the gas-liquid phase transition  beyond the Andrews critical point is of the {\it third order}.  
This explains why it is hard to observe the phase  liquid-gas transition  beyond the Andrews point, and 
clearly

\bigskip

{\sc Fourth}, physical intuition and the theory lead us to introduce asymmetry principle  of fluctuations, and 
the preferred transition mechanism.

\section{A Dynamic Model for  Gas-Liquid Transition}
\la{s8.1.1}

The classical and the simplest equation of state which can
exhibit many of the essential features of the gas-liquid phase
transition is the van der Waals equation:
\begin{equation}
\left(p + \frac{a}{v^2}\right) (v-b)=RT,\label{8.1}
\end{equation}
where $v$ is the molar  volume, $p$ is  the pressure, $T$ is the
temperature, $R$ is the universal gas constant, $b$ is  the revised
constant of inherent volume, and  $a$  is  the revised constant of attractive
force between molecules. If we adopt the molar density $\rho =1/v$
to replace $v$ in (\ref{8.1}), then the van der Waals equation
becomes
\begin{equation}
-(bp+RT)\rho +a\rho^2-ab\rho^3+p=0.\label{8.2}
\end{equation}

Now, we shall apply thermodynamic potentials to investigate the
gas-liquid phase transitions in $PVT$ systems, and we shall  see later that the
van der Waals equation can be derived as a Euler-Langrange equation for the minimizers of the Gibbs free energy for $PVT$ systems at gaseous states.

Consider an isothermal-isopiestic  process. The thermodynamic
potential is taken to be the Gibbs free energy. In this case, the
order parameters are the molar density $\rho$ and the  entropy density $S$, and
the control parameters are the pressure $p$ and temperature $T$. 
The general form of the  Gibbs free energy 
for $PVT$ systems  is given as 
 \begin{equation}
G(\rho ,S,T,p)   = \int_{\Omega}\Big[\frac{\mu_1}{2}|\nabla\rho
|^2+\frac{\mu_2}{2}|\nabla S|^2+g(\rho ,S,T,p)
  -ST-\alpha (\rho ,T,p)p   \Big]dx, \label{8.3}
 \end{equation}
where $g$ and $\alpha$ are differentiable with respect to  $\rho$ and $S$,
$\Omega\subset \R^3$ is the container, and $\alpha p $ is the
mechanical coupling term in the Gibbs free energy, which can be
expressed by
\begin{equation}
\alpha (\rho ,T,p)p=\rho p-\frac{1}{2}b\rho^2p, \label{8.4}
\end{equation}
where $b=b(T,p)$ depends continuously on $T$ and $p$. In fact, this mechanical coupling term should be $p$. In view of the van der Waals equation ~(\ref{8.2})  and the mathematical analysis based on the new dynamical transition theory, phenomenologically we need to adjust the term by adding a coefficient $\alpha$, leading 
to (\ref{8.4}) as the first two terms in  the Taylor expansion. Although the van der Waals equation works for gaseous sates only, by choosing the dependence of the coefficient $b$ on the temperature and the pressure, the energy applies to the liquid and solid states as well.  This is a very subtle term from the physical point of view to derive a feasible free energy. 

Based on both the  physical and mathematical considerations, 
we take the Taylor expansion of $g(\rho ,S,T,p)$ on $\rho$ and $S$ as follows
\begin{equation}
g=\frac{1}{2}\alpha_1\rho^2+\frac{1}{2}\beta_1 S^2+\beta_2S\rho^2-\frac{1}{3}\alpha_2\rho^3+\frac{1}{4}\alpha_3\rho^4,\label{8.5}
\end{equation}
where $\alpha_i$  $(1\leq i\leq 3)$, $\beta_1$  and $\beta_2$ depend
continuously on $T$ and $p$, and
\begin{equation}
\beta_1=\beta_1(T,p)>0, \qquad \alpha_i=\alpha_i(T,p)>0 \quad i=2,3.
\label{8.6}
\end{equation}

 In a $PVT$ system, the order parameter is $u=(\rho ,S)$,
$$\rho =\rho_1-\rho_0,\ \ \ \ S=S_1-S_0,$$
where $\rho_i$ and $S_i$  $(i=0,1)$ represent the density and entropy, 
$\rho_0,S_0$ are reference points near the coexistence curve of gas and liquid states. Hence the
conjugate variables of $\rho$ and $S$ are the pressure $p$ and
the temperature $T$. 
Thus, by the le Ch\^atlier principle, we derive from (\ref{8.3})-(\ref{8.5}) 
the following dynamic model for  a $PVT$ system: 
\begin{equation}
\left.
\begin{aligned} 
&\frac{\partial\rho}{\partial
t}=\mu_1\Delta\rho -(\alpha_1+bp)\rho
+\alpha_2\rho^2  -\alpha_3\rho^3-2\beta_2\rho S+p,\\
&\frac{\partial S}{\partial t}=\mu_2\Delta
S-\beta_1S-\beta_2\rho^2+T.
\end{aligned}
\right.\label{8.7}
\end{equation}

A physically meaningful boundary condition for the system is the Neumann boundary condition:
\begin{equation}
\frac{\partial\rho}{\partial n}   =0,\ \ \ \
\frac{\partial S}{\partial
n}  = 0 \qquad \text{ on } \partial\Omega. \label{8.8}
\end{equation}

An important special case for  $PVT$ systems is that the
pressure and temperature functions are homogeneous in $\Omega$. Thus we can
assume that $\rho$ and $S$ are independent of $x\in\Omega$,  and 
 the free energy (\ref{8.3}) with (\ref{8.4}) and
(\ref{8.5}) can be expressed as
\begin{equation}
 G(\rho,S,T,p)  = \frac{\alpha_1}{2}\rho^2+\frac{\beta_1}{2}S^2+\beta_2S\rho^2-
 \frac{\alpha_2}{3}\rho^3 +
\frac{\alpha_3}{4}\rho^4
+\frac{b\rho^2p}{2}-\rho p-ST.  \label{8.9}
\end{equation}

From (\ref{8.9}) we get the dynamical equations as
\begin{equation}
\left.
\begin{aligned} &\frac{d\rho}{dt}=-(\alpha_1+bp)\rho
+\alpha_2\rho^2-\alpha_3\rho^3-2\beta_2S\rho +p,\\
&\frac{dS}{dt}=-\beta_1S-\beta_2\rho^2+T.
\end{aligned}
\right.\label{8.10}
\end{equation}
Because $\beta_1>0$ for all $T$ and $p$, we can replace the second
equation of (\ref{8.10}) by 
\begin{equation}
S=\beta^{-1}_1(T-\beta_2\rho^2).
\end{equation}
Then, (\ref{8.10}) are equivalent to the following equation
\begin{equation} 
\frac{d\rho}{dt}=  -(\alpha_1+bp+2\beta^{-1}_1\beta_2T)\rho
+\alpha_2\rho^2 -(\alpha_3-2\beta^2_2\beta^{-1}_1)\rho^3+p.\label{8.11}
\end{equation}

It is clear that if  $\alpha_1=0, 2\beta^{-1}_1\beta_2=R,
\alpha_2=a, (\alpha_3-2\beta^2_2\beta^{-1}_1)=ab$,   then the steady
state equation of (\ref{8.11}) is referred to the van der Waals
equation.

We remark that(\ref{8.11}) can be considered as the dynamic version of the van der Waals equation, although we used the Landau mean field theory together with the le Ch\^atlier principle. The approach provides a much richer information. For example, the model (\ref{8.7}) can be used to study the heterogeneity of the system. 
In addition, the  model here can be used to study liquid-solid and gas-solid transitions as well, by choosing different parameters.

\section{Two New Physical Parameters  and the Andrews Critical Point}
In this section we use (\ref{8.11}) to derive two new physical parameters, which dictates the dynamic transition behavior near the Andrews critical points. 

Let $\rho_0$ be a steady state
solution of (\ref{8.11}) near  the Andrews point $C=(T_c, p_c)$. We take the transformation
$$\rho =\rho_0+\rho^{\prime}.$$
Then equation (\ref{8.11}) becomes (drop the prime)
\begin{equation}
\frac{d\rho}{dt}=\lambda\rho +a_2\rho^2-ab \rho^3.\label{8.12}
\end{equation}
where
\begin{equation}
\begin{aligned}
&\lambda
=2a \rho_0-3ab \rho^2_0-\alpha_1-bp-R T,\\
&a_2=a (1-3b \rho_0).
\end{aligned}
\end{equation}
where $\alpha_1$  is close to zero. Here we emphasize that $\rho_0$  and $(\lambda, a_2)$ are all functions of the control parameters $(T, p)$.

These are two important physical parameters, which are used to fully characterize the dynamic behavior of gas-liquid transition near the Andrews point. In fact, from the derivation of the model, we obtain immediately the following physical meaning of these two parameters:
\begin{equation}\label{2p}
\lambda(T, p)= \frac{d^2 G}{d \rho^2} \big|_{\rho=\rho_0}, \qquad 
a_2(T, p)= \frac12 \frac{d^3 G}{d \rho^3} \big|_{\rho=\rho_0},
\end{equation}
where $\rho_0=\rho_0(T, p)$  is the equilibrium state.

In   the $PT$-plane, near  the Andrews point $C=(T_c, p_c)$,  the critical parameter equation
$$\lambda =\lambda (T,p)=0 \qquad  \text{in}\ \ \ \ |T-T_c|<\delta
,\ \ \ \ |p-p_c|<\delta$$ 
for some $\delta >0$,
defines  a continuous function $T=\phi (p)$, such that
\begin{equation}
\lambda
\left\{
\begin{aligned} 
& <0  &&  \text{ if }  T>\phi (p),\\
& =0  &&  \text{ if } T=\phi (p),\\
& >0  &&  \text{ if } T<\phi (p).
\end{aligned}
\right.\label{8.13}
\end{equation}
Equivalently, this is called the principle of exchange of stabilities, which, as we have shown in \cite{MW08a}, is the necessary and sufficient condition for the gas-liquid phase transition.

One important component of our theory is that the Andrews critical point is determined by the system of equations
\begin{equation}
\begin{aligned}
& \lambda=0, \\
& a_2 = 0, \\
& -(bp+ RT) \rho_0 + a \rho_0^2 -ab \rho_0^3 + p =0.
\end{aligned}
\end{equation}
Here the first equation is the critical parameter equation, the second equation, as we shall see below, determines the switching point where the phase transition switches types, and the last equation is the van der Waals equation, which is also the steady state equation of the dynamic model.

Then by a direct computation, it is easy to see that the critical point $C$ is given by 
\begin{equation}\label{critical-point}
(\rho_c, T_c, p_c)=\left(\frac{1}{3b}, \frac{a}{27 b^2}, \frac{8a}{27bR}\right).
\end{equation}
This is in agreement with the classical work by van der Waals. Here we obtain the Andrews point using a dynamic approach.

\section{Theory of the Andrews Critical Point}
We now explain the gas-liquid transition near the Andrews critical point $C$. 

\medskip

{\sc First}, we have shown in (\ref{critical-point})   that at the equilibrium point $\rho_0$, the two curves given by $\lambda(p, T)=0$  and $a_2(p, T)=0$ interact exactly at the critical point $C$ as shown in Figure~\ref{f8.9}, and the
curve segment $AB$ of $\lambda =0$ is divided into two parts $AC$
and $CB$ by the point $C$ such that
\begin{align*}
& a_2(T,p)>0\ \ \ \ \text{for}\ \ \ \ (T,p)\in AC,\\
& a_2(T,p)<0\ \ \ \ \text{for}\ \ \ \ (T,p)\in CB.
\end{align*}
\begin{SCfigure}[25][t]
  \centering
  \includegraphics[width=0.35\textwidth]{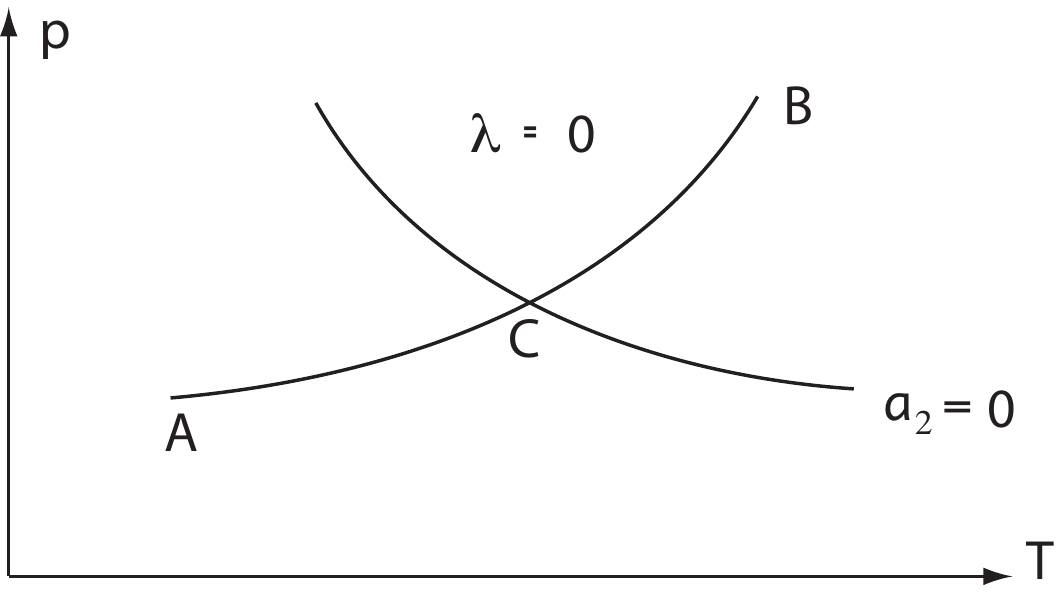}
  \caption{The point $C=(T_c,p_c)$ is the Andrews critical
point.}\la{f8.9}
 \end{SCfigure}
Here the curve $AC$ is  the classical gas-liquid co-existence curve.

\medskip

{\sc Second}, on the curve $AC$, excluding the critical point $C$, $a_2(T, p) >0$. The phase transition of the system is a mixed type if we take a pass crossing the $AC$; see  Figure~\ref{f8.4}. In Figure~\ref{f8.4}, $\rho$  is the deviation from the basic gaseous state $\rho_0$.   We now consider different states given in Figure~\ref{f8.4}(a):

\begin{enumerate}
\begin{figure}[hbt] 
  \centering
  \includegraphics[width=0.35\textwidth]{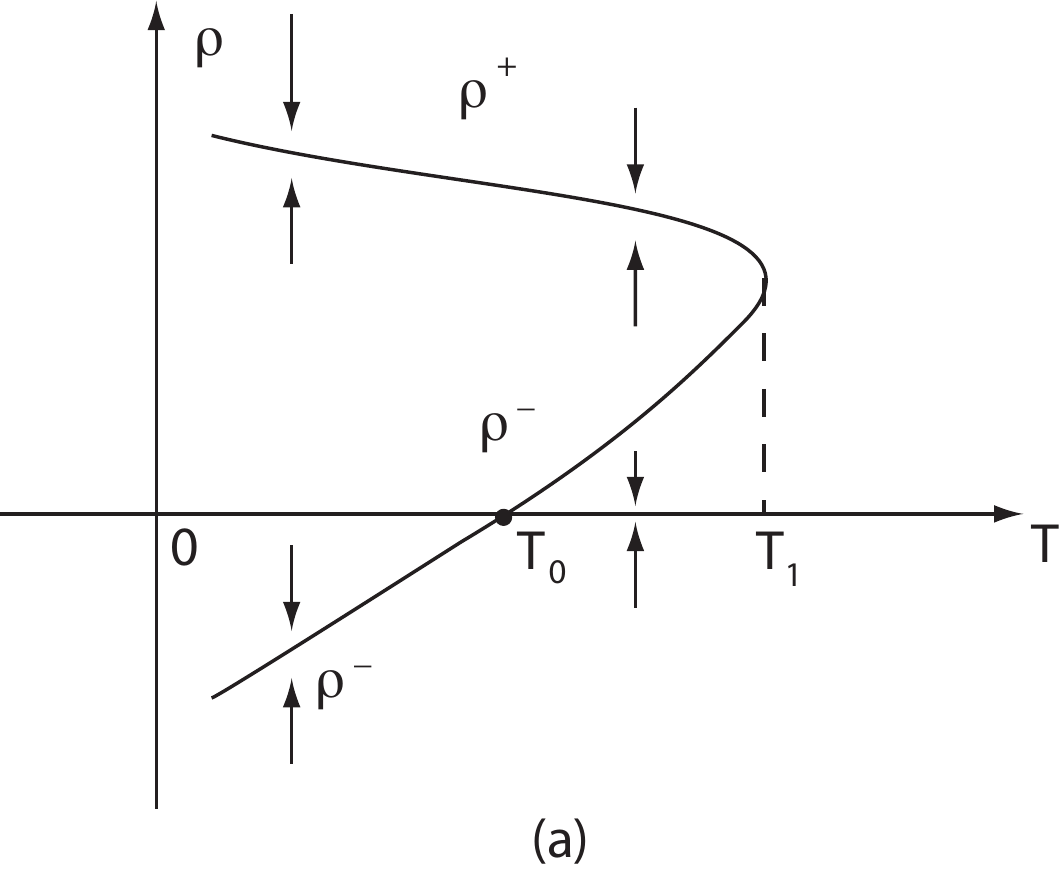} 
  \includegraphics[width=0.35\textwidth]{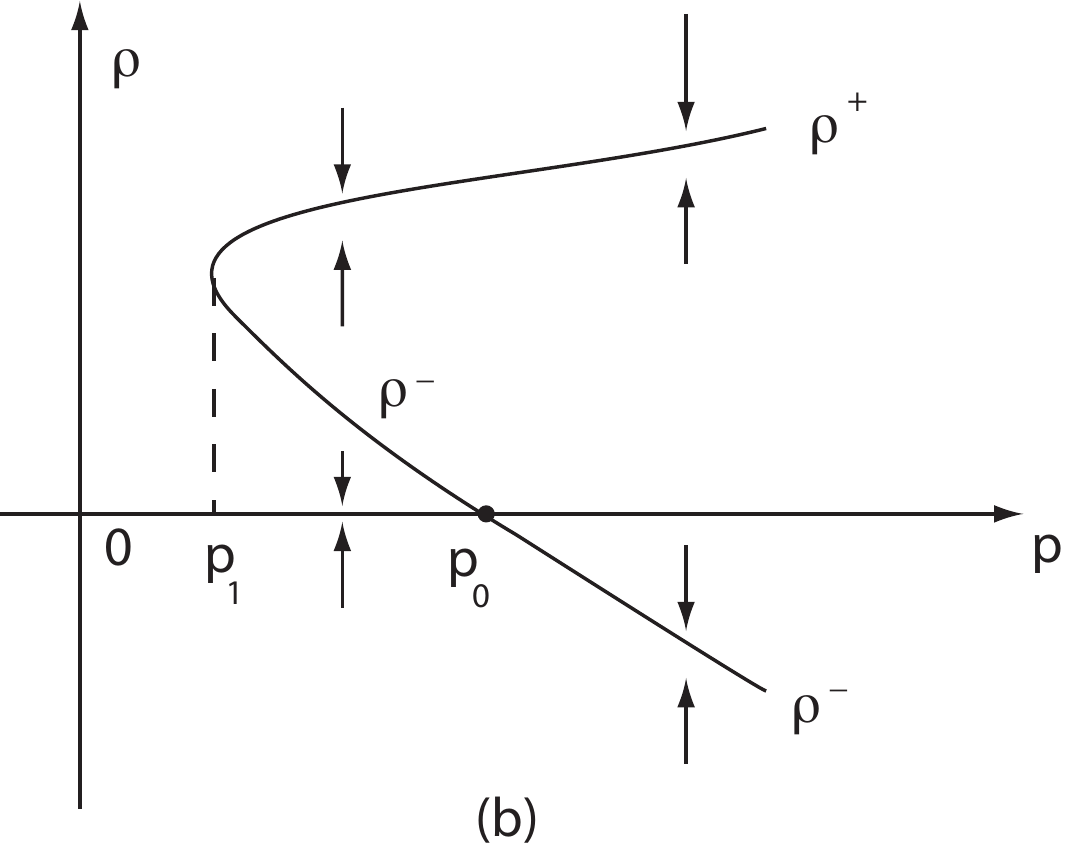}
  \caption{Type-III (mixed) transition for $a_2>0$: (a) Phase diagram for fixed pressure, and (b) phase diagram for fixed temperature.}\la{f8.4}
 \end{figure}

\item For $T>T_1$, the gaseous state $\rho_0$, corresponding to $\rho=0$ in the figure,  is stable. This is the only stable physical state in this temperature range, and the system is in the gaseous state.

\item For $T_0 < T < T_1$, there are two  metastable states given by gaseous phase $\rho_0$ 
and the liquid phase $\rho_0 + \rho^+$.

\item For $T < T_0$, there are three states: the unstable basic gaseous state $\rho_0$, and  the 
two metastable states: $\rho_0 + \rho^-$  and $\rho_0 + \rho^+$. 
One important component of our theory is that the only physical phase here is the liquid phase represented by 
metastable state: $\rho_0 + \rho^+$. Although, mathematically speaking, the gaseous state 
$\rho_0 + \rho^-$  is also metastable, it does not appear in nature. The only possible explanation for this exclusion is   the asymmetry principle of fluctuations, to be further explored in the next section.

\item Hence we have shown that as we lower the temperature, the system undergoes 
a {\it first order} transition from a gaseous state  to a liquid state with an abrupt change in
density. In fact, there is an energy gap
between the gaseous and liquid states:
$$
\Delta E=G(\rho_0+\rho^+)-G(\rho_0)\sim - \frac{\lambda}{4} (\rho^+)^2 - \frac{a_2}{12} (\rho^+)^3<0\ \ \ \ \text{for }\ T<T_1;
$$
see \cite{MW08a}.
This energy gap $|\Delta E|$ stands for a latent heat, and $\Delta
E<0$ shows that the transition from a gaseous state to a liquid state is an
isothermal exothermal process, and from a liquid state  to gaseous state is an
isothermal endothermal process.

\end{enumerate}

{\sc Third}, at the critical point $C$, we have $a_2=0$. Then the dynamic transition is as shown in 
Figure~\ref{f8.3}; see \cite{MW08a} for the detailed mathematical analysis leading to this phase diagram:

\begin{enumerate}
\item As in the previous case, for $T> T_0$, the only physical state is given by the gaseous phase $\rho_0$, corresponding to zero deviation shown in the figure.

\item As the temperature $T$ is lowered crossing $T_0$, the gaseous state losses its stability, leading to two metastable states: one is the liquid phase $\rho_0 + \rho^+$, and the other is the gaseous phase $\rho_0 + \rho^-$. Again, the gaseous phase $\rho_0 + \rho^-$  does not appear, and the asymmetry principle of fluctuations is valid in this situation as well.

\item The phase transition here is of the second order, as the energy is continuous at $T_0$. In fact, the energy for the transition liquid state is given by 
$$
G( \rho_0 + \rho^+ )=G(\rho_0)-\frac{\alpha^2}{4a_3}(T-T_c)^2\ \ \ \
\text{for}\ T<T_c,
$$ 
for some $\alpha >0$.
Hence the difference of the heat capacity at $T=T_c$ is
$$
\Delta C=-T_c\frac{\partial^2}{\partial
T^2}\left(G(\Phi^+(T_c^+))-G(\rho_0)\right)=\frac{\alpha^2}{2a_3}T_c>0.
$$
Namely the heat capacity has a finite jump at $T=T_c$, therefore
the transition at $T=T_c$ is of  the second order.

\end{enumerate}

\begin{figure}[hbt]
  \centering
  \includegraphics[width=0.35\textwidth]{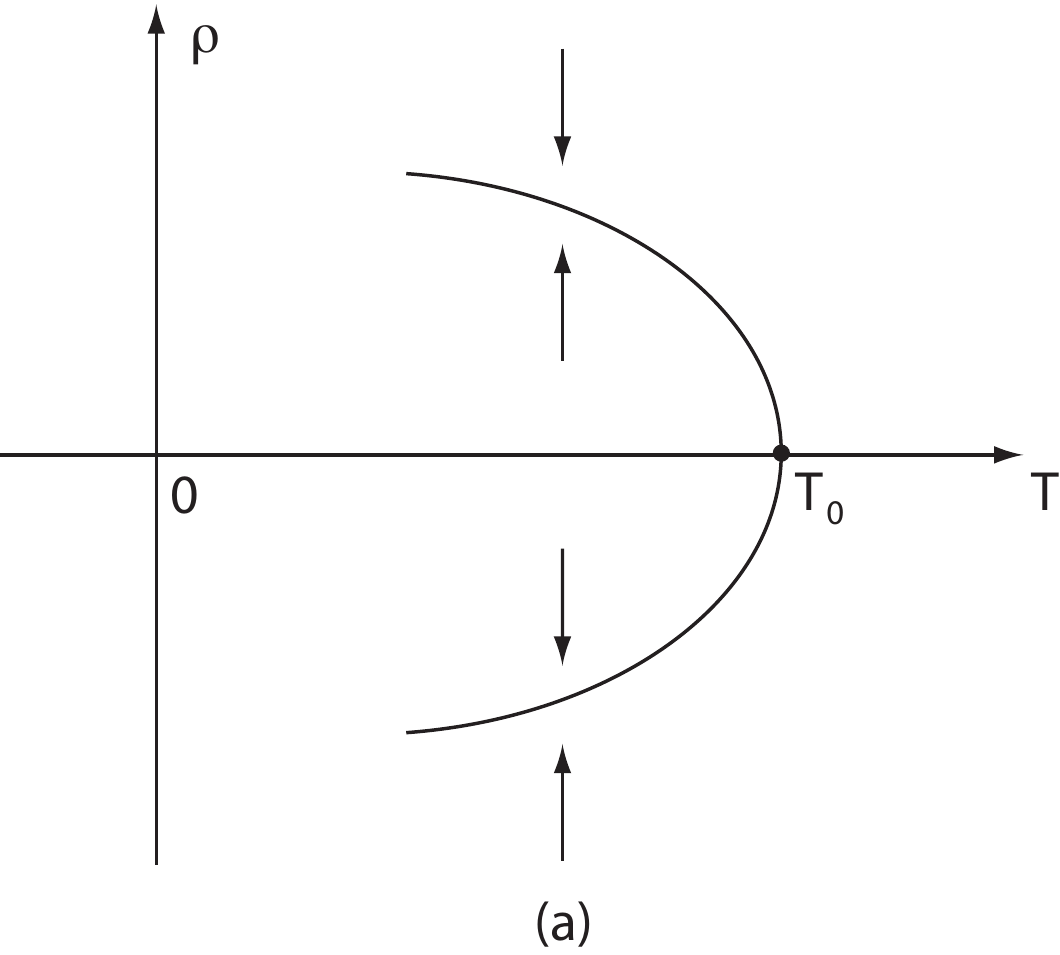}
  \includegraphics[width=0.35\textwidth]{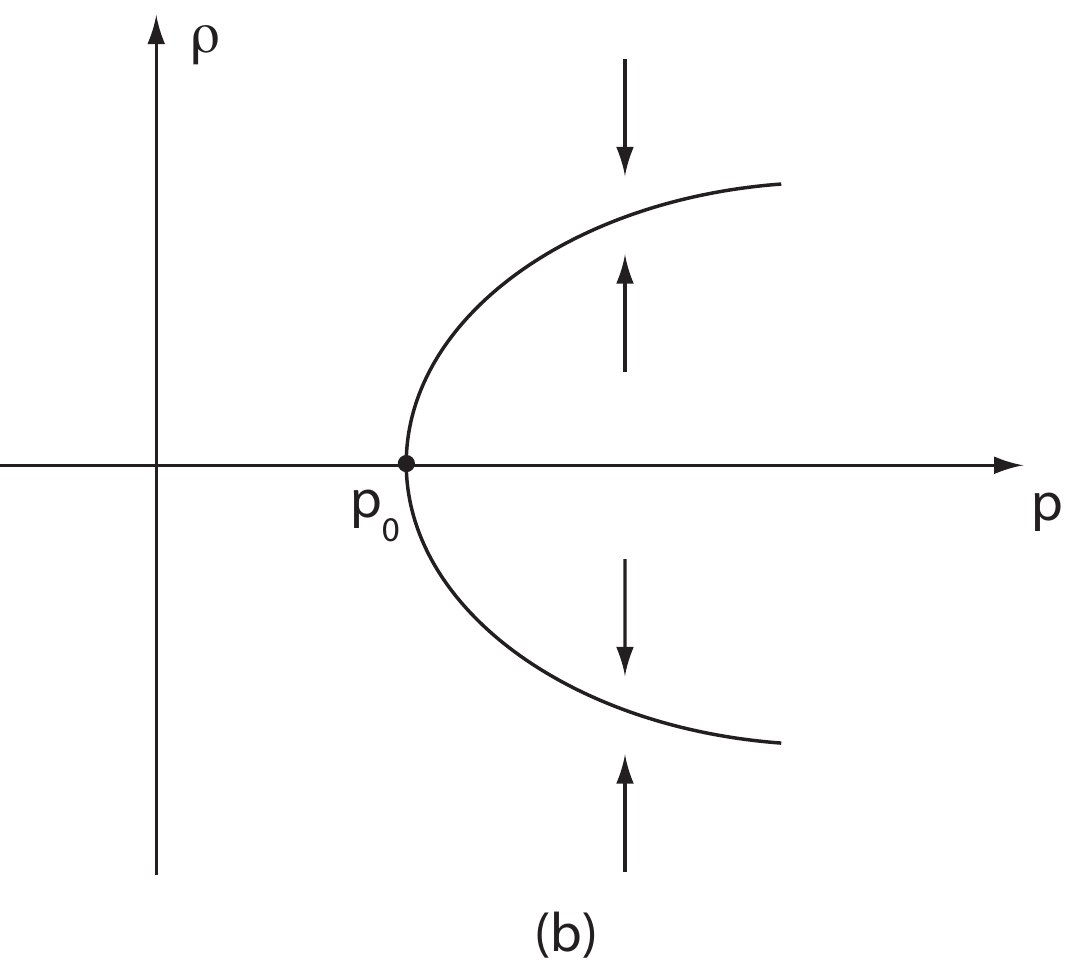}
  \caption{Continuous transition for the case where
$a_2=0.$}\la{f8.3}
 \end{figure}

{\sc Fourth}, on the curve $BC$, $a_2(T, p) < 0$, and the phase transition diagram is 
given by Figure~\ref{f8.5}:

\begin{enumerate}

\item For $T> T_1$, the system is in the gaseous phase, which is stable. 

\item For $T_0 < T < T_1$, there are two metastable gaseous states given by $\rho_0$ and $\rho_0 + \rho^-$. As before, although it is metastable, the gaseous state $\rho_0 + \rho^-$ does not appear. 

\item For $T< T_0$, the gaseous phase $\rho_0$  losses its stability, and the system undergoes a dynamic transition to the metastable  liquid state $\rho_0 + \rho^+$. Mathematically,  the gaseous state $\rho_0 + \rho^-$  is also metastable. However, it  does not appear either, due to the asymmetry principle of fluctuations.

\item The dynamic transition in this case is of the third order. In fact, we have 
$$G(\rho_0 + \rho^+ )=G(\rho_0(T))-\frac{\alpha^3}{6|a_2|^2}(T_0-T)^3+o(|T_0-T|^3) \text{ for } T < T_0.$$
Namely, the free energy  is continuously
differentiable up to the second order at $T=T_0$, and the
transition is of the third order. It implies that as $(T_0,p_0)\in CB$
the third-order  transition at $(T_0,p_0)$ can not be observed by
physical experiments.

\end{enumerate}

\begin{figure}[hbt]
  \centering
  \includegraphics[width=0.35\textwidth]{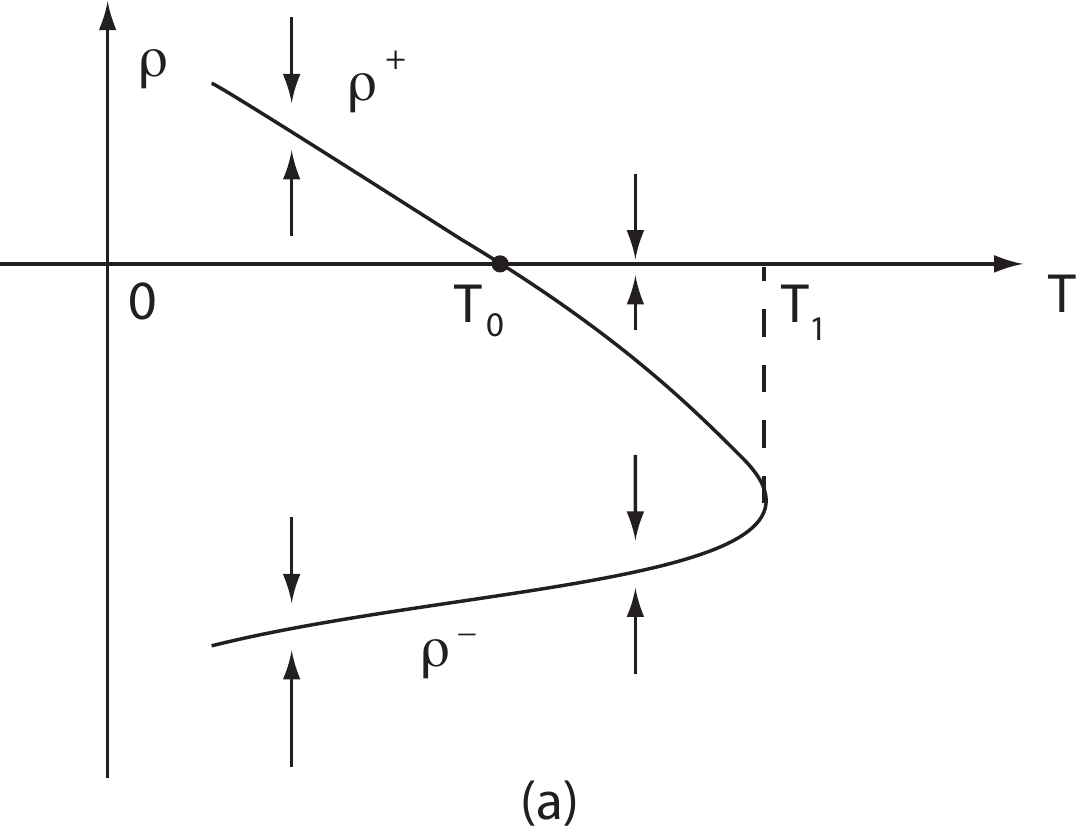}
  \includegraphics[width=0.35\textwidth]{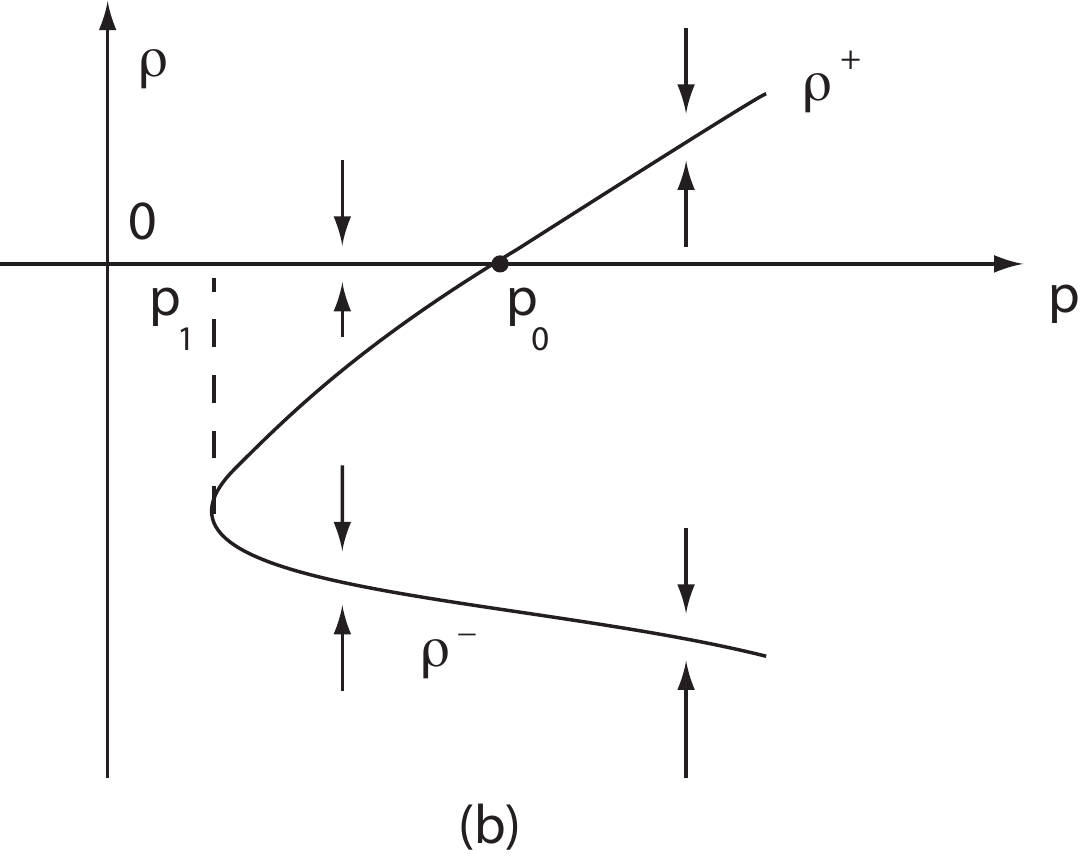}
  \caption{Type-III (mixed) transition for $a_2<0$.}\la{f8.5}
 \end{figure}

In summary, we have obtained a precise characterization of the phase transition behavior near the Andrews point, and have derived precisely the nature of the Andrews critical point. In particular, we have shown the following:

\begin{enumerate}

\item The transition is first order before the critical point, second-order at the critical point, and third order after the critical point. 

\item  The curve $\lambda(T, p)=0$ always defines the gas-liquid co-existence curve in both sides of the critical point $C$. We note that  in the classical theory, the co-existence curve terminates at the critical point, and with our theory, we are able to determine the gas-liquid transition behavior and the co-existence curve beyond the critical point.
\end{enumerate}

\section{Asymmetry Principle of Fluctuations and the Preferred Transition Mechanism} 
We have shown in the last section that in all three cases (both sides of the critical point and at the critical point),  the metastable state $\rho_0 + \rho^-$ does not appear. 
Hence the only possible physical explanation is the asymmetry of the fluctuations. In fact,  for the
ferromagnetic systems we also see this asymmetry of fluctuations  \cite{MW08b}. This observation leads to the following important principle:

\bigskip
\noindent
{\bf  Physical Principle} (Asymmetry of Fluctuations). 
{\it 
 The symmetry of fluctuations for general thermodynamic systems may not
be universally true. In other words, in some systems with
multi-equilibrium metastable states, the fluctuations near a critical point
occur only in one basin of attraction of some equilibrium states,
which are the ones that can be physically observed.}

\bigskip

An alternate explanation of this principle is  related to phase  transitions in certain preferred direction in a given thermodynamic system, which we call preferred  transition mechanism. We conjecture that this mechanism 
is universal as well. Here we use this mechanism to explain the asymmetry principle of fluctuations in the  gas-liquid transition, and, in return,  to explain  the meaning of the preferred  transition mechanism.

In the gas-liquid transition as the temperature is lowered, the system prefers phase transitions to denser phase. 
This can be considered as one aspect of the preferred transition mechanism. 
\begin{figure}[hbt]
  \centering
  \includegraphics[width=0.35\textwidth]{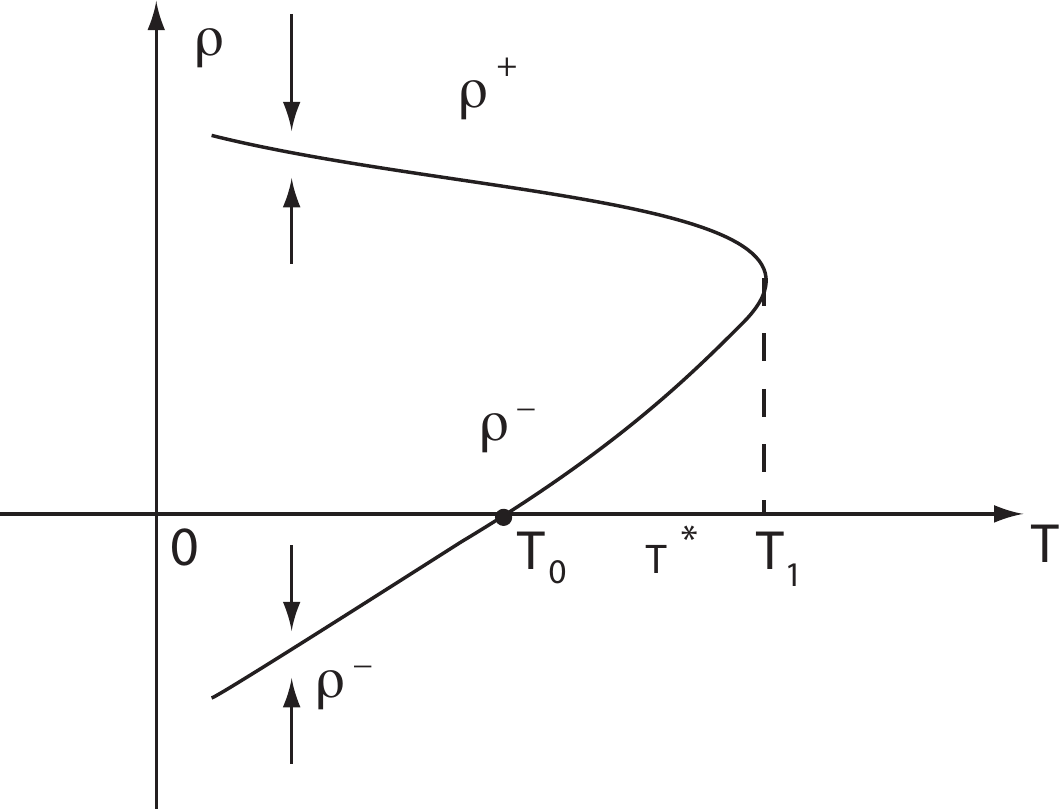} 
  \caption{Preferred transition mechanism for $a_2>0$ with  fixed pressure.}\la{f8.4-1}
 \end{figure}

Another important aspect of the mechanism is the preferred  transition at a critical point $T^\ast$ as shown in Figure~\ref{f8.4-1}, which is reproduced from Figure~\ref{f8.4}(a). This critical point is between $T_0$ and $T_1$, and for water under one atmospheric  pressure, $T^*$  is 100 $^o$C. For $T^* < T < T_1$,  the  liquid state state $\rho_0 + \rho^+$ is called superheated liquid, and for $T_0 < T < T^*$, the gaseous state $\rho_0$  is called supercooled gas. The preferred  transition mechanism consists of the following:

\begin{enumerate}

\item As temperature decreases, the system  is forced to  undergo a first-order transition at $T= T^*$, from the gas state $\rho_0$ to the liquid state $\rho_0 + \rho^+$.

\item As the temperature increases, the system is forced to undergo a first-order liquid $\rho_0+ \rho^+$ to gas 
$\rho_0$  transition as the same critical point $T=T^*$.

\item The supercooled gas and superheated liquid can rarely occur. Physically, this is related to the hysteresis phenomena.

\end{enumerate}

\bibliographystyle{siam}

\begin{thebibliography}{1}

\bibitem{MW08a}
{\sc T.~Ma and S.~Wang}, {\em Dynamic phase transition theory in {PVT}
  systems}, Indiana University Mathematics Journal, 57:6 (2008),
  pp.~2861--2889.

\bibitem{MW08b}
\leavevmode\vrule height 2pt depth -1.6pt width 23pt, {\em Dynamic transitions
  for ferromagnetism}, Journal of Mathematical Physics, 49:053506 (2008),
  pp.~1--18.

\bibitem{reichl}
{\sc L.~E. Reichl}, {\em A modern course in statistical physics}, A
  Wiley-Interscience Publication, John Wiley \& Sons Inc., New York,
  second~ed., 1998.

\bibitem{stanley}
{\sc H.~E. Stanley}, {\em Introduction to Phase Transitions and Critical
  Phenomena}, Oxford University Press, New York and Oxford, 1971.

\end{thebibliography}

\end{document}